# Protocol Design and Stability/Delay Analysis of Half-Duplex Buffered Cognitive Relay Systems

Yan Chen, Vincent K. N. Lau, Shunqing Zhang, and Peiliang Qiu

*Abstract*—In this paper, we quantify the benefits of employing relay station in large-coverage cognitive radio systems which opportunistically access the licensed spectrum of some small-coverage primary systems scattered inside. Through analytical study, we show that even a simple decode-and-forward (SDF) relay, which can hold only one packet, offers significant *path-loss gain* in terms of the spatial transmission opportunities and link reliability. However, such scheme fails to capture the spatial-temporal burstiness of the primary activities, that is, when either the source-relay (SR) link or relay-destination (RD) link is blocked by the primary activities, the cognitive spectrum access has to stop. To overcome this obstacle, we further propose buffered decode-and-forward (BDF) protocol. By exploiting the infinitely long buffer at the relay, the blockage time on either SR or RD link is saved for cognitive spectrum access. The *buffer gain* is shown analytically to improve the stability region and average end-to-end delay performance of the cognitive relay system.

## I. INTRODUCTION

The core idea behind the cognitive transmission is to let the secondary users (SUs) exploit the under-utilized spectrum holes left by the primary communication systems [1]–[3], either in temporal, frequency or spatial domain, without interfering the regular transmissions of the primary users (PU). One key issue of the cognitive radio (CR) system is on the efficiency of spectrum sharing with the PU system. Direct transmission, which demands large transmit power, ends up with small opportunity of access and hence low spectrum sharing efficiency. As such, *CR combined with relay station (RS)*, referred to as *cognitive relay system* (CRS), appears as an attractive solution to boost the spectrum sharing efficiency.

The majority of the existing works on CRS focused on the physical layer aspects of the problem [4]–[6]. For example, a distributed algorithm for channel access and power control was proposed for cognitive multi-hop relays in [6], and a channel selection policy for multi-hop cognitive mesh network was considered by [5]. When *delay-sensitive* applications are considered, other performance measures such as the *stability region* and the *average end-to-end packet delay* become critical. In [7], the authors analyzed the delay of a cognitive relay assisted multi-access network, however, they did not consider

Manuscript received 1 Dec 2008; revised 7 May 2009 and 3 Sep 2009; accepted 11 Nov 2009. The associate editor coordinating the review of this paper and approving it for publication was Prof. Ashutosh Sabharwal.

Yan Chen and Peiliang Qiu (e-mail: qiupl418,qiupl@zju.edu.cn) are with Institute of Information and Communication Engineering, Zhejiang University, Hangzhou, China. Dr. Yan Chen is now working in Huawei Technologies, Shanghai, China. V. K. N. Lau and S. Zhang (e-mail: eeknlau,eezsq@ee.ust.hk) is with Department of Electronic and Computer Engineering, Hong Kong University of Science and Technology (HKUST), Clear Water Bay, Kowloon, Hong Kong.

This paper is funded by RGC 615407.

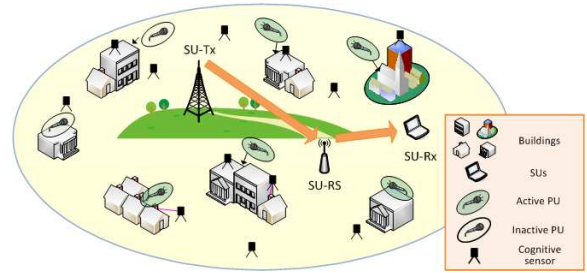

Fig. 1. An example of large-coverage CRS opportunistically access the licensed spectrum of some randomly distributed small-coverage PU systems.

the impact of PU activities and dynamic spectrum-sharing. Moreover, in the existing works, the coverage of the PU system is assumed to be much larger than the SU's, so the spatial burstiness of the primary traffic and its impact on the CRS have not been fully investigated.

In this paper, we try to shed a light on the protocol design of CRS that addresses the above issues. We are interested in the scenario where the CRS coverage is much larger than the PU coverage and try to design compatible cognitive relay protocols that could effectively deal with the uncertainty of PU locations and the spatial burstiness of PU activities. In addition, we study the stability region and the average end-to-end delay of the CRS under different relaying protocols, which are critical for delay-sensitive applications. We shall show that the introduction of a cognitive relay provides two levels of potential gains. In particular, a conventional decode-and-forward (DF) relay under a simple DF (SDF) protocol provides increased spatial transmission opportunities and enhanced link reliability, which is referred to as *path-loss gain*. On top of it, by enabling buffering at the relay, the blockage time on either the source-relay or relay-destination link can be saved to further increase spectrum access opportunities and reduce the end-to-end delay. This is referred to as *buffer gain* and is shown via the analysis of our proposed buffered DF (BDF) protocol. We derive the closed-form expressions of the stability region and the average end-to-end delay for each protocol and quantify the two types of gains based on the derived results.

## II. SYSTEM MODEL

We consider a scenario where the coverage of the SU system is much larger than that of the PU systems. One example of such CR network is the WRAN system covering a suburb college town or rural areas, whose cell radius ranges from 2-10 km or even larger. While the PU systems inside are Part74 devices (wireless microphone), whose transmission ranges are about 100-200 m. So the transmission between a pair of SU nodes may affect multiple PU systems simultaneously.

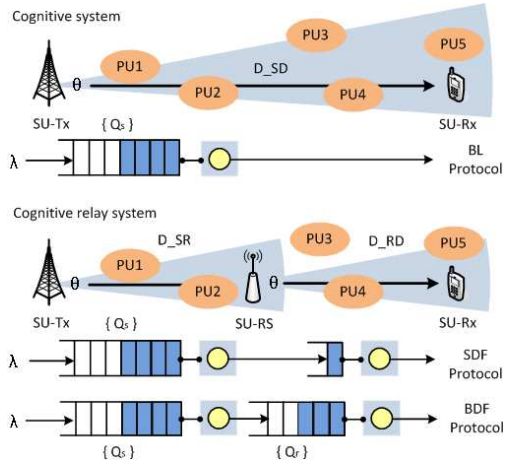

Fig. 2. Illustrative diagram of the CRS under three protocols.

### A. Assumptions for SU Transmission

We consider a CRS with a SU transmitter (SU-Tx), a SU receiver (SU-Rx), and a half-duplex cognitive RS (SU-RS), as shown in Fig. 2. The links between SU-Tx and SU-Rx, SU-Tx and SU-RS, SU-RS and SU-Rx are referred to as SD, SR and RD links, respectively. In order to reduce the interference region caused by SU transmission, we assume that antenna arrays are equipped at both SU-Tx and SU-RS for beamforming [1] with beamwidth $\theta$ and transmit antenna gain $G_t$. Since SU-RS works in half-duplex mode, it uses the antenna array to obtain receiving antenna gain $G_r$. At SU-Rx, however, only one omnidirectional antenna is available. Further assume the transmission time of the CRS is slotted. In any slot, using power $P_{ij}$ to transmit signal $X$ (with unit signal energy) on link $ij$, the received signal at $j$ is

$$Y_j = \begin{cases} a(P_{ij})h_{ij}\sqrt{P_{ij}L_{ij}G_t}X + I_j + Z_j, & i = S, R, j = D \\ a(P_{ij})h_{ij}\sqrt{P_{ij}L_{ij}G_tG_r}X + I_j + Z_j, & i = S, j = R \end{cases}$$

respectively, where $a(P_{ij})$ is an indicator variable and is a function of $P_{ij}$. It indicates whether the transmission from $i$ to $j$ using power $P_{ij}$ is blocked by any PU, $a(P_{ij}) = 0$ indicates the transmission is blocked and $a(P_{ij}) = 1$ otherwise. $h_{ij}$ stands for the channel fading coefficient, which is assumed to be flat Rayleigh so that the power gain on the link $ij$, i.e. $H_{ij} = |h_{ij}|^2$, is exponentially distributed with parameter 1. $H_{ij}$ is assumed to be quasi-static within a slot but identically and independently distributed (i.i.d) between different slots. $L_{ij} = \kappa_0 \cdot D_{ij}^{-\alpha}$ is the large-scale path-loss between $i$ and $j$, where $D_{ij}$ is the distance between $i$ and $j$, $\kappa_0$ and $\alpha$ are path-loss coefficient and exponent, respectively. $I_j$ stands for the sum signals received from all neighboring active PUs at node $j$. Since the PU's coverage is much smaller and we assume SU-Rx is not inside the coverage of any active PU, then we can treat $I_j$ as white noise with power $\mathbb{E}[I_j] = \sigma_I^2$, $\forall j$. $Z_j$

---

is the white Gaussian noise at receiver $j$, i.e. $Z_j \sim \mathcal{N}(0, \sigma_j^2)$ and we assume $\sigma_j^2 = \sigma^2$, $\forall j$. Further define the instantaneous received signal-to-interference-and-noise ratio (SINR) on link $ij$ as $\gamma_{ij} = a(P_{ij}) \cdot C_0 \cdot \frac{P_{ij}}{D_{ij}^\alpha} \cdot H_{ij}$, $j = D$ and $\gamma_{SR} = a(P_{SR}) \cdot C_0 G_r \cdot \frac{P_{SR}}{D_{SR}^\alpha} \cdot H_{SR}$, respectively, where $C_0 = \frac{\kappa_0 G_t}{\sigma^2 + \sigma_I^2}$ is a constant independent of the transmit power. Moreover, we assume the maximum transmit power constraint for SU-Tx and SU-RS are both $P_{max}$ respectively.

### B. Assumptions for PU Distribution and Activities

We assume the PUs are uniformly (randomly) distributed on the two-dimensional plane with density $\rho$. In any given slot, each PU can be either active (ON) (with probability $\pi_1$) or inactive (OFF) (with probability $\pi_0 = 1 - \pi_1$). By the indicator variable $a(P_{ij})$, we have related the impact of PU activities to received SINRs on link $ij$. When transmitting with power $P_{ij}$, the average interference-to-noise ratio (INR) received by a PU $D_{SP}$ distance away is $\bar{\gamma}_{SP}(P_{ij}, D_{SP}) = C_0 \cdot \frac{P_{ij}}{\sigma_I^2 D_{SP}^\alpha}$. When the INR is higher than threshold $\gamma_{th}$, the PU transmission would be interfered. Setting $\bar{\gamma}_{SP}(P_{ij}, D_{SP}^*) = \gamma_{th}$, we can find the radius of the maximum interference region as $D_{SP}^* = \left( \frac{C_0}{\sigma_I^2 \gamma_{th}} \cdot P_{ij} \right)^{1/\alpha}$. Since the directional beam with beamwidth $\theta$ rad can be approximated by a sector with angle $\frac{\theta}{2\pi}$, the area of the interference region can be approximated by the area of that sector with radius $D_{SP}^*$, which is $A_{SP}(P_{ij}) = \pi D_{SP}^{*2} \cdot \frac{\theta}{2\pi}$. According to the uniform distribution assumption, the average number of PUs in the interference region can thus be calculated as $N(P_{ij}) = \rho \cdot A_{SP}(P_{ij})$. The SU transmission on link $ij$ is "blocked" if any of the $N(P_{ij})$ PUs is active and $a(P_{ij}) = 0$. Recall $S_k$ is the activity state of the $k$-th PU in the region and the probability that the SU transmission with power $P_{ij}$ would not be blocked is

$$\Pr\{a(P_{ij}) = 1\} = \pi_0^{N(P_{ij})} = \exp\left\{ -C_1(\rho, \pi_0) \cdot P_{ij}^{2/\alpha} \right\}$$

where $C_1(\rho, \pi_0) = \frac{\rho\theta}{2} \left( \frac{C_0}{\sigma_I^2 \gamma_{th}} \right)^{2/\alpha} \ln \frac{1}{\pi_0} > 0$ is a constant independent of the transmit power $P_{ij}$ but directly related to the PU distribution density $\rho$ and activity intensity $\pi_0$.

### C. Probability of Successful Transmission over a Link

We assume the data of the SU system is encapsulated into small packets with $M$ bits each. For each time slot, at most one packet can be transmitted which requires channel capacity larger than $M$. Using Shannon formula, the channel capacity between node $i$ and $j$ can be expressed as $\Phi_{ij} = B \log_2(1 + \gamma_{ij})$, where $B$ is the bandwidth of the channel and $\gamma_{ij}$ is the SINR at receiver $j$. Thus, the probability of successful transmission of a packet over link $ij$, denoted as $p_{ij}^{succ}$, is

$$p_{ij}^{succ}(P_{ij}) = \Pr\{B \log_2(1 + \gamma_{ij}) > M\}$$
$$= \Pr\left\{ H_{ij} > \frac{(2^{M/B} - 1)D_{ij}^\alpha}{C_0 P_{ij}} \right\} \cdot \Pr\{a(P_{ij}) = 1\}$$
$$\overset{(1)}{=} \exp\left\{ -C_1(\rho, \pi_0) \cdot P_{ij}^{2/\alpha} - C_2(M/B, D_{ij}) \cdot P_{ij}^{-1} \right\}, \quad (1)$$

---

[1] Beamforming increases the ave. Rx SNR at the SU-Rx but the instantaneous Rx SNR still follows Rayleigh fading due to the local scattering cluster.

for $i = S, R, j = D$, where $C_2(M/B, D_{ij}) = \frac{(2^{M/B}-1)}{C_0} \cdot D_{ij}^{\alpha} > 0$ is a constant independent of $P_{ij}$ but directly related to the ratio of packet size over the channel bandwidth $M/B$ and the distance between the two ends of the link $D_{ij}$. Step (1) is from equation (1). For the case $j = R$,

$$p_{SR}^{succ}(P_{SR}) = \exp\left\{-\frac{C_1(\rho, \pi_0)}{P_{SR}^{-2/\alpha}} - \frac{C_2(\frac{M}{B}, D_{SR})}{G_r P_{SR}}\right\}. \quad (2)$$

*Property 1:* For a given set of $d_{ij}$, $R$, $\rho$, and $\pi_0$, there exists a unique transmit power $P_{ij}^{opt} > 0$ that maximize the probability of successful transmission over link $ij$, i.e. $\exists P_{ij}^{opt}$, s.t. $P_{ij}^{opt} = \arg\max_{P_{ij}} p_{ij}^{succ}(P_{ij})$. The optimal transmit power for SD, SR and RD links (in the interference limited case, i.e. $P_{ij}^{opt} < P_{max}$) are $(i = S, R)$

$$P_{iD}^{opt} = \left(\frac{\alpha C_2(\frac{M}{B}, D_{iD})}{2C_1(\rho, \pi_0)}\right)^{\frac{\alpha}{\alpha+2}} \quad P_{SR}^{opt} = \left(\frac{\alpha C_2(\frac{M}{B}, D_{SR})}{2G_r C_1(\rho, \pi_0)}\right)^{\frac{\alpha}{\alpha+2}} \quad (3)$$

respectively, and the maximized probabilities of successful transmission on SD, RD and SR links are $(i = S, R)$

$$p_{iD}^{succ,opt} = \exp\left\{-C_3 C_1^{\frac{\alpha}{\alpha+2}}(\rho, \pi_0) C_2^{\frac{2}{\alpha+2}}(\frac{M}{B}, D_{iD})\right\} \quad (4)$$

$$p_{SR}^{succ,opt} = \exp\left\{-\frac{C_3}{G_r^{\frac{2}{\alpha+2}}} C_1^{\frac{\alpha}{\alpha+2}}(\rho, \pi_0) C_2^{\frac{2}{\alpha+2}}(\frac{M}{B}, D_{SR})\right\} \quad (5)$$

where $C_3 = \left(1 + \frac{\alpha}{2}\right)\left(\frac{\alpha}{2}\right)^{-\frac{\alpha}{\alpha+2}}$ is a constant related to $\alpha$.

*Remark 1:* For fixed transmit power $P_{ij}$, the impact of PU activity on the SU transmission is like a good/bad fading process, with states $a(P_{ij}) = 1$ (good) and $a(P_{ij}) = 0$ (bad). However, it is different from a traditional channel fading process ($\rho = 0$, $\pi_0 = 1$) when the transmit power can be adjusted. Specifically, to combat a traditional channel fading, increasing the transmit power always helps to increase $p_{ij}^{succ}$, while with random PU distribution and activities, $p_{ij}^{succ}(P_{ij})$ is not a monotonic increasing function of $P_{ij}$. As shown in the Property 1, there exist a unique $P_{ij}^{opt}$ that maximize $p_{ij}^{succ}(P_t)$ for a specific link $ij$.

## III. PROTOCOL DESCRIPTIONS

In each protocol, an infinitely long buffer is assumed at SU-Tx and the first-in-first-out rule is applied. Recall that the transmission time is slotted and the packet transmission starts at the beginning of a slot. In each slot, only one packet can be transmitted. When transmitting, SU-Tx and SU-RS are supposed to use the optimal transmit power to maximize the probability of successful transmission over a link.

**Baseline (BL) Protocol:** In the BL protocol, SU-Tx transmits a packet *directly* to SU-Rx if the SD link is not blocked. The packet is removed from SU-Tx's queue when an acknowledge (ACK) message is received from SU-Rx[2].

**Simple Decode-and-Forward (SDF) Protocol:** In the SDF protocol, SU-RS can successfully receive a packet from SU-Tx is the SR link is not blocked. It decodes the received packet and forwards to the SU-Rx as a conventional DF relay does. The SU-RS can not receive new packet from SU-Tx before the currently holding packet has been successfully forwarded to the SU-Rx. A packet can be removed from SU-Tx's queue if an ACK is received from SU-Rx.

**Buffered Decode-and-Forward (BDF) Protocol:** In BDF protocol, infinitely long buffer is also assumed at SU-Tx. SU-Tx transmits a packet to SU-RS only when SU-RS is not transmitting on the RD link and the SR link is not blocked. The packet is removed from SU-Tx's queue if an ACK is received from SU-RS. Higher transmission priority is assumed at SU-RS such that it transmits to SU-Rx whenever the RD link is not blocked. Moreover, we assume SU-RS has the channel state information (CSI) of the RD link so that the channel outage time of the RD link can be further saved for SR link transmission. The packet is removed from SU-RS's relay queue when an ACK is received from SU-Rx.

## IV. ANALYSIS OF STABILITY/DELAY PERFORMANCE

### A. Queue Dynamics and Stability/Delay Definitions

**Queue Dynamics:** We adopt a similar model used in [7], [8] to depict the buffer dynamics in a slotted system. Let $Q_S(t), t = m\tau, m = 0, 1, \ldots$ denote the queue length at the SU-Tx observed at the end of slot $t$. It evolves as $Q_S(t) = (Q_S(t-1) - Y_S(t))^+ + X_S(t), \forall t$. In the equation, $X_S(t)$ represents the number of packet arrivals in slot $t$ (cannot be transmitted in the same slot), which is assumed to be a Bernoulli process with mean $\mathbb{E}[X_S(t)] = \lambda$, i.e. $X_S(t)$ only takes value 0 or 1 with probability $1 - \lambda$ and $\lambda$, respectively. $Y_S(t)$ denotes the number of packets that depart from SU-Tx in slot $t$. According to the protocols, $Y_S(t)$ also takes value from $\{0, 1\}$, depending on the states of PU activities, channel fading, and the interaction with the queue dynamics at SU-RS. Since $Q_S(t+1)$ only depends on $Q_S(t)$ and $X_S(t), Y_S(t)$ is either 0 or 1, $\{Q_S\}$ is a discrete time Markov Chain and its state transitions only happen between neighboring states, i.e. $\{Q_S\}$ is a discrete time birth-death process (DTBDP). For $n \geq 0$, let $\lambda_S^n$ be the state transition probabilities from $Q_S(t) = n$ to $Q_S(t+1) = n+1$ and $\mu_S^n$ the state transition probabilities from $Q_S(t) = n$ to $Q_S(t+1) = n - 1$ for $n \geq 1$. Similarly, the evolution for the queue at SU-RS can be defined as $Q_R(t) = (Q_R(t-1) - Y_R(t))^+ + X_R(t), \forall t$. Note the arrival process $X_R(t)$ depends on the departure process of the source queue (i.e. $Y_S(t)$) and the half-duplex constraint makes the interaction between $Q_S(t)$ and $Q_R(t)$ complicated. $\{Q_R\}$ is also a DTBDP, whose state transition probabilities can be defined in a similar manner as $\lambda_R^n$ and $\mu_R^n$, respectively.

**Stability of a CRS:** A queue $Q_i$ is *stable* if and only if $\lim_{t \to \infty} \Pr\{Q_i(t) = 0\} > 0$ [8], for $i = S, R$. In the BL and SDF protocols, when the queue at SU-Tx is stable, the whole system is stable, but in the BDF protocol, system stability requires both the queue at SU-Tx and the queue at SU-RS to be stable simultaneously. Denote by $\lambda^*$ the stability region of the CRS in terms of the maximum exogenous arrival rate, which has unit "packet/slot".

---

**End-to-End Delay of a CRS:** The *end-to-end delay* of a packet in the CRS is the time from a packet arrives at the queue of SU-Tx till the packet reaches SU-Rx. Little's theorem [9] enables the study from the angle of average queue length. Given the exogenous arrival rate $\lambda$, the average end-to-end delay for the three protocols can be defined as (unit: slots):

$$\overline{W}_{BL} = \frac{1}{\lambda} \lim_{T \to \infty} \frac{\sum_{t=1}^{T} Q_S(t)}{T} = \frac{\mathbb{E}[Q_S]}{\lambda}, \qquad (6)$$

$$\overline{W}_{SDF/BDF} = \frac{1}{\lambda} \lim_{T \to \infty} \frac{\sum_{t=1}^{T} Q_S(t)}{T} + \frac{1}{\lambda} \lim_{T \to \infty} \frac{\sum_{t=1}^{T} Q_R(t)}{T}$$
$$= \frac{\mathbb{E}[Q_S] + \mathbb{E}[Q_R]}{\lambda} \qquad (7)$$

where $\mathbb{E}$ is taken w.r.t the steady distribution of queue length.

### B. Stability/Delay Analysis for the BL protocol

In the BL protocol, only $\{Q_S\}$ is involved. The queue length increases by one if a new packet arrives and no packet has been successfully transmitted, and decrease by one if an existing packet is successfully transmitted and no new packet has arrived, which implies $\lambda_S^{n=0} = \lambda$ and $\lambda_S^{n \geq 1} = \lambda \cdot (1 - p_{SD}^{succ,opt})$, and $\mu_S^{n \geq 1} = (1 - \lambda) \cdot p_{SD}^{succ,opt}$. We derive the stable distribution of $\{Q_S\}$ by solving the detailed balance equation related to the DTBDP. Define $q_S^n$ as $\Pr[Q_S = n]$, we have

$$q_S^0 = \left(1 + \sum_{n=1}^{\infty} \prod_{k=0}^{n-1} \frac{\lambda_S^k}{\mu_S^{k+1}}\right)^{-1} = 1 - \frac{\lambda}{p_{SD}^{succ,opt}}, \qquad (8)$$

$$q_S^n = q_S^0 \prod_{k=0}^{n-1} \frac{\lambda_S^k}{\mu_S^{k+1}} = \frac{q_S^0}{1 - p_{SD}^{succ,opt}} \left(\frac{\lambda(1 - p_{SD}^{succ,opt})}{(1-\lambda) \cdot p_{SD}^{succ,opt}}\right)^n \quad (9)$$

System stability requires $q_S^0 > 0$, which implies $\lambda^* = p_{SD}^{succ,opt}$. Furthermore, the end-to-end delay can be calculated as according to (6). Theorem 1 summarized the above results.

*Theorem 1:* The stability region and the average end-to-end delay under the BL protocol are

$$\lambda_{BL}^* = p_{SD}^{succ,opt}, \qquad \overline{W}_{BL} = (1-\lambda) / (p_{SD}^{succ,opt} - \lambda) (10)$$

### C. Stability/Delay Analysis for the SDF/BDF protocol

In the following, we shall first analyze a general case of SU-RS having buffer length $L$.

**General Case:** Similar as in the BL protocol, the decrease in $Q_S$ implies a packet has been successfully transmitted to SU-RS and no new packet arrives. However, under the SDF or BDF protocol, the departure process of $Q_S(t)$ is related to the dynamics of $Q_R(t)$. The departure of a packet implies that the following three events must be true simultaneously: 1) The queue length at SU-RS is not $L$, denoted as $\mathcal{E}\{Q_R \neq L\}$; 2) SU-RS is not transmitting on the RD link, denoted as $\mathcal{E}\{$RD link idle$\}$; 3) The SR link is able to support the packet transmission, denoted as $\mathcal{E}\{$SR link successful$\}$. Therefore[3],

$$\mu_S^{n \geq 1} = (1 - \lambda) \Pr\left(\mathcal{E}\{\text{SR link successful}\}\right)$$
$$\times \Pr\left(\mathcal{E}\{\text{RD link idle}\} \big| \mathcal{E}\{Q_R \neq L\}\right) \Pr\left(\mathcal{E}\{Q_R \neq L\}\right)$$
$$= (1 - \lambda) \left[q_R^0 + (1 - q_R^0 - q_R^L)(1 - p_{RD}^{succ})\right] p_{SR}^{succ,opt}$$
$$\lambda_S^{n \geq 1} = \lambda \left\{1 - \left[q_R^0 + (1 - q_R^0 - q_R^L)(1 - p_{RD}^{succ})\right] p_{SR}^{succ,opt}\right\}$$

and $\lambda_S^{n=0} = \lambda$. Using the same method as in obtaining (8), the steady state probability of $Q_S = 0$ is

$$q_S^0 = 1 - \frac{\lambda}{\left[q_R^0 + (1 - q_R^0 - q_R^L)(1 - p_{RD}^{succ})\right] p_{SR}^{succ,opt}} (11)$$

which implies the stability region is

$$\lambda^* = \left[q_R^0 + (1 - q_R^0 - q_R^L)(1 - p_{RD}^{succ})\right] p_{SR}^{succ,opt}. \quad (12)$$

For $\{Q_R\}$, due to the half-duplex constraint, packet arrival and packet departure would not happen in the same slot. As the increase of $Q_R$ has lower priority than its decrease. So the probability that $Q_R$ decreases by one equals the probability of successful transmission on the RD link. Thus, the state transition probabilities of $\{Q_R\}$ are

$$\lambda_R^{n=0} = (1 - q_S^0) p_{SR}^{succ,opt},$$
$$\lambda_R^{1 \leq n \leq L-1} = (1 - q_S^0) p_{SR}^{succ}(1 - p_{RD}^{succ}),$$
$$\mu_R^{1 \leq n \leq L} = p_{RD}^{succ,opt}.$$

So the steady state probability of $Q_R = 0$ is

$$q_R^0 = \left(1 + \frac{\sum_{k=1}^{L} \left(\frac{(1-q_S^0) p_{SR}^{succ,opt}(1 - p_{RD}^{succ,opt})}{p_{RD}^{succ,opt}}\right)^k}{1 - p_{RD}^{succ,opt}}\right)^{-1} (13)$$

The steady state probability of $Q_R = k$ can be further obtained via similar calculations used in (9) as

$$q_R^k = \frac{q_R^0}{p_{RD}^{succ,opt}} \left((1 - q_S^0) p_{SR}^{succ,opt}(1 - p_{RD}^{succ,opt}) / p_{RD}^{succ,opt}\right)^k$$

Since $q_R^L$ is function of $q_R^0$, solving the combined equations of (11) and (13), we can get the solutions for $q_S^0$ and $q_R^0$, which further lead to the results of the stability region $\lambda^*$ and the average end-to-end delay [4]. We now apply the general results derived above to the SDF and BDF protocols, respectively.

**Special Case I - SDF Protocol:** In this case, $L = 1$ and $\{Q_R\}$ reduces to a two-state Markov chain with $q_R^L = 1 - q_R^0$. Therefore, (11) and (13) reduce to

$$q_S^0 = 1 - \lambda / q_R^0 p_{SR}^{succ,opt},$$
$$q_R^0 = p_{RD}^{succ,opt} / \left((1 - q_S^0) p_{SR}^{succ,opt} + p_{RD}^{succ,opt}\right).$$

---

[3] Here we approximate $\Pr\{Q_R(t) = m \mid Q_S(t) = n\}$ as $\Pr\{Q_R(t) = m\}$, which is asymptotically tight for large $L$ case (e.g. the BDF case) but is an upper bound on $\mu_S^n$ in the SDF case. As a result, the final expression of the stability region (end-to-end delay) of the SDF case is an upper bound (lower bound), which is verified by the simulation in Section V.

[4] There is no closed-form expressions for $q_R^0$ and $q_S^0$ in the general case, which involves finding the roots of an polynomial to a power of $L$. However, for the two special cases $L = 1$ and $L = \infty$, the polynomials reduce to quadratic forms and closed-form solutions are ready to get.

Solving the two equations, the steady state probabilities are

$$q_S = 1 - \lambda/(p_{SR}^{succ,opt} - p_{SR}^{succ,opt}\lambda/p_{RD}^{succ,opt}), \quad (14)$$

$$q_R = 1 - \lambda/p_{RD}^{succ,opt}, \quad (15)$$

respectively. The stability region can be obtained from the requirement $q_S^0 > 0$ and the average end-to-end delay can be calculated according to (7). Theorem 2 summarizes the results.

*Theorem 2:* The stability region and the average end-to-end delay for the SDF protocol are

$$\lambda_{SDF}^* = \frac{p_{RD}^{succ,opt} p_{SR}^{succ,opt}}{p_{RD}^{succ,opt} + p_{SR}^{succ,opt}} \quad (16)$$

$$\overline{W}_{SDF} = \frac{1-\lambda}{p_{SR}^{succ,opt} - \frac{p_{SR}^{succ,opt}}{p_{RD}^{succ,opt}}\lambda - \lambda} + \frac{1}{p_{RD}^{succ,opt}}. \quad (17)$$

**Special Case II - BDF Protocol:** In this case, $L = \infty$ and then for a stable system $q_R^L = 0$. Correspondingly, we have

$$q_S^0 = 1 - \frac{\lambda}{[1-(1-q_R^0)p_{RD}^{succ,opt}]p_{SR}^{succ,opt}},$$

$$q_R^0 = 1 - \frac{(1-q_S^0)p_{SR}^{succ,opt}}{p_{RD}^{succ,opt}\left[1+(1-q_S^0)p_{SR}^{succ,opt}\right]}.$$

Solving the two equations, we get

$$q_S^0 = 1 - \lambda/\left[(1-\lambda)p_{SR}^{succ,opt}\right], \quad q_R^0 = 1 - \lambda/p_{RD}^{succ,opt} \quad (18)$$

respectively. The stability region can be obtained from satisfying the requirements of both $q_S^0 > 0$ and $q_R^0 > 0$, and the average end-to-end delay can be calculated according to (7). Theorem 3 summarizes the above results.

*Theorem 3:* The stability region and the average end-to-end delay for the BDF protocol are

$$\lambda_{BDF}^* = \min\left\{\frac{p_{SR}^{succ,opt}}{1+p_{SR}^{succ,opt}}, p_{RD}^{succ,opt}\right\}, \quad (19)$$

$$\overline{W}_{BDF} = \frac{1-\lambda}{(1-\lambda)p_{SR}^{succ,opt} - \lambda} + \frac{1-\lambda}{p_{RD}^{succ,opt} - \lambda}. \quad (20)$$

## V. ANALYSIS OF PATH-LOSS GAIN AND BUFFER GAIN

### A. Analysis of Path-loss Gain

Path-loss gain is referred to as the gain that a CRS obtains from the reduction in path-loss over the BL system. We shall use the metric called "throughput per Watt" to illustrate this gain, which is defined as the average throughput (in bits/slot) of delivering a single packet from SU-Tx to SU-Rx divided by the transmit power needed. In this case, the SDF and BDF protocols behave alike, so we only compare the SDF protocol with the BL protocol.

To transmit a single packet with $R$ bits on SD, SR, and RD links take $1/p_{SD}^{succ,opt}$, $1/p_{SR}^{succ,opt}$ and $1/p_{RD}^{succ,opt}$ time slots, respectively, so the average throughput for the BL and SDF protocols are $T_{BL} = R \cdot p_{SD}^{succ,opt}$ and $T_{SDF} = \frac{R}{1/p_{SR}^{succ,opt}+1/p_{RD}^{succ,opt}} = \frac{R \cdot p_{SR}^{succ,opt} p_{RD}^{succ,opt}}{p_{SR}^{succ,opt}+p_{RD}^{succ,opt}}$, respectively. While the transmit power needed for the two protocols are $P_{SD}^{opt}$ and $P_{SR}^{opt} + P_{RD}^{opt}$, respectively. Further define $\Delta T = T_{SDF}/T_{BL} - 1$ and $\Delta P = (P_{SR}^{opt} + P_{RD}^{opt})/P_{SD}^{opt} - 1$ as the throughput gain and power gain of the SDF protocol over the BL protocol, then the path-loss gain in throughput per Watt is thus given by $\Delta T_P = \frac{1+\Delta T}{1+\Delta P} - 1$. Fig. 3 depicts the path-loss gain of a CRS over the BL system as a function of SU-RS's location. Here we assume SU-Tx and SU-Rx are fixed at $(0,0)$ and $(2,0)$ km while SU-RS is at $(D_{SR},0)$. It can be shown that setting $P_{SR}^{opt}(D_{SR}) = P_{RD}^{opt}(D_{SD} - D_{SR})$ achieves the peak of the path-loss gain, which implies $D_{SR}^* = (1-C_4)D_{SD}, D_{RD}^* = C_4 D_{SD}$, where $C_4 = \frac{G_r^{-1/\alpha}}{1+G_r^{-1/\alpha}}$.

*Property 2:* The path-loss gain achieves its peak when SU-RS is located at $(D_{SR}^*,0)$, i.e.,

$$\Delta T_P^* = \frac{1+\Delta T}{1+\Delta P} - 1 = \frac{\frac{1}{2}\left(p_{SD}^{succ,opt}\right)^{C_4^{\frac{2\alpha}{\alpha+2}}-1}}{G_r^{-\frac{\alpha}{\alpha+2}}(1-C_4)^{\frac{\alpha^2}{\alpha+2}} + C_4^{\frac{\alpha^2}{\alpha+2}}} - 1.$$

Given such location of SU-RS, the condition for having positive path-loss gain is given by

$$p_{SD}^{succ,opt} < \left(2G_r^{-\frac{\alpha}{\alpha+2}}(1-C_4)^{\frac{\alpha^2}{\alpha+2}} + 2C_4^{\frac{\alpha^2}{\alpha+2}}\right)^{\frac{1}{C_4^{\frac{2\alpha}{\alpha+2}}-1}}.$$

### B. Analysis of Buffer Gain

The buffer gain is referred to the gain that comes from enabling the buffering capability at SU-RS. We shall use the results derived in section IV to illustrate this gain. In particular, we shall show the relative increase in the stability region and the reduction in the average end-to-end delay under the BDF protocol (infinite buffer) compared with the SDF protocol (single packet storage). The buffer gain in the stability region and in the average end-to-end delay are defined as $\Delta\lambda^* = \lambda_{BDF}^*/\lambda_{SDF}^* - 1$ and $\Delta\overline{W} = 1 - \overline{W}_{BDF}/\overline{W}_{SDF}$, respectively.

*Property 3:* Given SU-RS located at $(D_{SR}^*,0)$, and let $\zeta \overset{\Delta}{=} p_{RD}^{succ,opt}|_{D_{RD}=D_{RD}^*} = \left(p_{SD}^{succ,opt}\right)^{C_4^{\frac{2\alpha}{\alpha+2}}}$, the buffer gain in stability region and in average end-to-end delay are

$$\Delta\lambda^* = \frac{1-\zeta}{1+\zeta} > 0, \qquad \Delta\overline{W} = \frac{\frac{\lambda^2}{1-\lambda}(1-\zeta)}{(\zeta-\lambda)(\zeta-\frac{\lambda}{1-\lambda})} > 0 \quad (21)$$

The value of $\zeta$ in (21) depends on the system parameters such as $\alpha$ and $G_r$ as well as PU distribution density $\rho$ and PU activity intensity $\pi_0$. Fig. 4 depicts how the buffer gain given by equation (21) varies with PU activity intensity $\pi_0$ and PU distribution density. From the figure we see that the buffer gain increases when the cognitive environment is more unfavorable (e.g. larger intensity of PU activity and higher density of PU distribution), which verifies that with the buffering capability enabled at the cognitive relay, a CRS can better adjust to the environment and more efficiently deal with the spatial burstiness of the PU activities.

### C. Numerical Discussions

In this subsection, we shall discuss the end-to-end delay performance under the three protocols and verify our analytical results via Monte Carlo simulations. The packet size $M = 16$ kbits and the channel bandwidth is $B = 16$ kHz. Other parameters are set as follows: $\kappa_0 = 1$, $\alpha = 4$, $\sigma^2 = \sigma_I^2 = 1$, $G_t = G_r = 2$, $\theta = \pi/3$ rad and $\gamma_{th} = 1$. Fig. 5 depicts the the performance of the average end-to-end delay with increasing PU distribution density, while Fig. 6 shows the trend that the average end-to-end delay varies with decreasing PU activity intensity. We can see that the simulation results matches well with our analytical results. Moreover, the BDF protocol always achieves larger stability region and smaller delay than the SDF protocol, as claimed in Property 3.

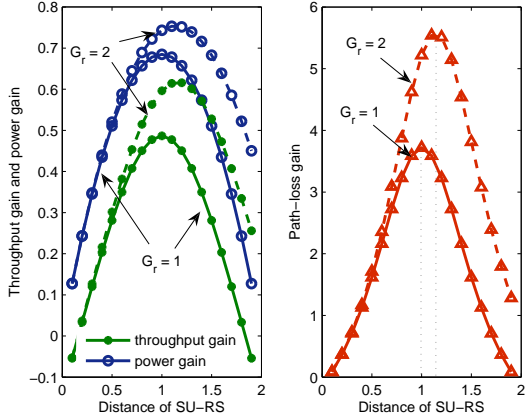

Fig. 3. Path-loss gain of CRS over BL system with different distances of SU-RS. $D_{SD} = 2$ km, $\rho = 2$ per km$^2$, $\pi_0 = 0.8$.

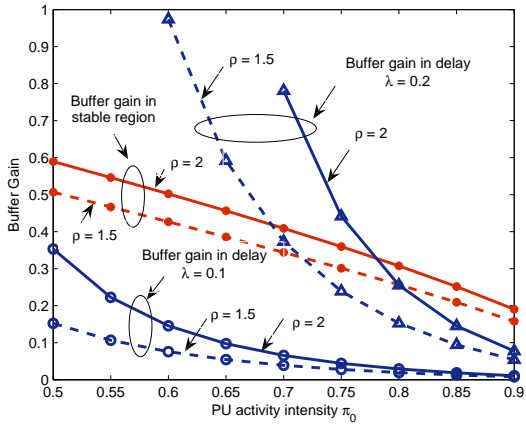

Fig. 4. Buffer gain in both stability region and average end-to-end delay of CRS under the BDF protocol over that under the SDF protocol with different PU activity intensity $\pi_0$. $\rho = 2$ per km$^2$, $D_{SD} = 2$ km, and SU-RS is fixed at $(D_{SR}^*, 0) = ((1 - C_4)D_{SD}, 0) = (1.086, 0)$ km.

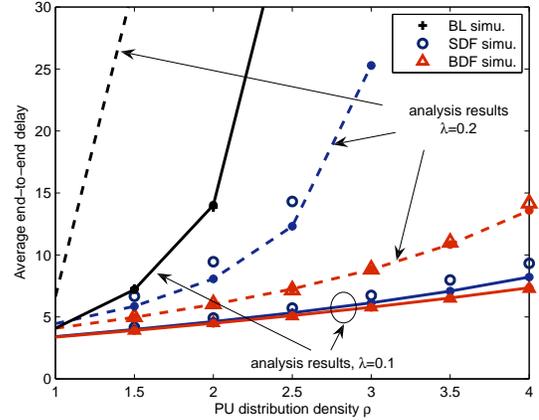

Fig. 5. The average end-to-end delay under the three protocols with different PU distribution density $\rho$ per km$^2$. $D_{SD} = 2$ km and SU-RS is fixed at $(D_{SR}^*, 0) = ((1 - C_4)D_{SD}, 0) = (1.086, 0)$ km, $\pi_0 = 0.8$.

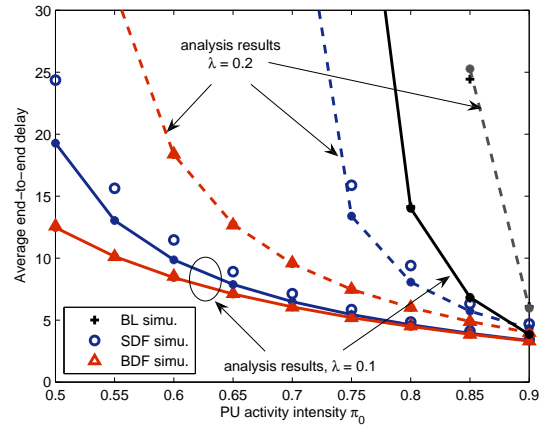

Fig. 6. The average end-to-end delay under the three protocols with different PU activity intensity $\pi_0$. $D_{SD} = 2$ km and SU-RS is fixed at $(D_{SR}^*, 0) = ((1 - C_4)D_{SD}, 0) = (1.086, 0)$ km, $\rho = 2$ per km$^2$.

## VI. CONCLUSION

We have shown two levels of gains provided by a CRS, i.e. *path-loss gain* and *buffer gain*. The path-loss gain substantially increases the spectrum access opportunities by reducing the transmit power while the buffering capability at the relay further saves the blockage time of either the SR or RD link and reduces the end-to-end delay to a larger extent. We emphasize the importance of exploiting relay buffers to deal with the uncertainty of PU activities, which is an intrinsic issue associated with large-coverage cognitive systems.

## REFERENCES

[1] J. Mitola and G. Q. Maguire, "Cognitive radio: making software radios more personal," *IEEE Personal Commun. Mag.*, vol. 6, no. 4, pp. 13 – 18, Aug. 1999.

[2] S. Haykin, "Cognitive radio: brain-empowered wireless communications," *IEEE J. Sel. Areas Commun.*, vol. 23, no. 2, pp. 201 – 220, Feb. 2005.

[3] Q. Zhao, L. Tong, A. Swami, and Y. Chen, "Decentralized cognitive mac for opportunistic spectrum access in ad hoc networks: A pomdp framework," *IEEE J. Sel. Areas Commun.*, vol. 25, no. 3, pp. 589 – 600, Apr. 2007.

[4] K. Lee and A. Yener, "Outage performance of cognitive wireless relay networks," in *Proc. IEEE GLOBECOM '06*, Nov. 2006.

[5] M. Sharma, A. Sahoo, and K. D. Nayak, "Channel selection under interference temperature model in multi-hop cognitive mesh networks," in *Proc. IEEE DySPAN '07*, Apr. 2007, pp. 133 –136.

[6] Y. T. Hou, Y. Shi, and H. D. Sherali, "Spectrum sharing for multi-hop networking with cognitive radios," *IEEE J. Sel. Areas Commun.*, vol. 26, no. 1, pp. 146 – 155, Jan. 2008.

[7] A. K. Sadek, K. J. R. Liu, and A. Ephremides, "Cognitive multiple access via cooperation: Protocol design and performance analysis," *IEEE Trans. Inf. Theory*, vol. 53, no. 10, 3677 – 3696, Oct. 2007.

[8] R. R. Rao and A. Ephremides, "On the stability of interacting queues in a multiple-access system," *IEEE Trans. Inf. Theory*, vol. 34, pp. 918 – 930, Sept 1988.

[9]  L. Kleinrock, *Queueing Systems Volume I: Theory*.  John Wiley & Sons, 1975.

[10]  [Online]. Available: http://arxiv.org/abs/0809.0533